\documentclass[aps,prx,reprint,groupedaddress,superscriptaddress,footnoteinbib]{revtex4-2}
\usepackage{amsmath,amssymb}
\usepackage{ bbold }
\usepackage{upgreek}
\usepackage{dsfont}
\usepackage{graphicx, epstopdf} 
\usepackage{amsfonts}
\usepackage{graphicx,xcolor}
\usepackage{amsmath, bbm}
\usepackage{natbib}
\usepackage{braket}
\usepackage[utf8x]{inputenc}
\usepackage[ngerman, english]{babel}
\usepackage[T1]{fontenc}
\usepackage{mathtools}
\usepackage{hyperref}
\usepackage{soul}

\begin{document}


\title{Enantiomer superpositions from matter-wave interference of chiral molecules}

\author{Benjamin A. Stickler}
\email{benjamin.stickler@uni-due.de}
\affiliation{Faculty of Physics, University of Duisburg-Essen, 47048 Duisburg, Germany}
\affiliation{QOLS, Blackett Laboratory, Imperial College London, London SW7 2AZ, United Kingdom}

\author{Mira Diekmann}
\affiliation{Fachbereich Chemie, Philipps-Universität Marburg, 35032 Marburg, Germany}

\author{Robert Berger}
\affiliation{Fachbereich Chemie, Philipps-Universität Marburg, 35032 Marburg, Germany}

\author{Daqing Wang}
\email{daqing.wang@uni-kassel.de}
\affiliation{Experimentalphysik I, Universit\"{a}t Kassel, 34132 Kassel, Germany}

\begin{abstract}
Molecular matter-wave interferometry enables novel strategies for manipulating the internal mechanical motion of complex molecules. Here, we show how chiral molecules can be prepared in a quantum superposition of two enantiomers by far-field matter-wave diffraction and how the resulting tunnelling dynamics can be observed. We determine the impact of ro-vibrational phase averaging and propose a setup for sensing enantiomer-dependent forces, parity-violating weak interactions, and environment-induced superselection of handedness, as suggested to resolve Hund's paradox. Using ab-initio tunnelling calculations, we identify [4]-helicene derivatives as promising candidates to implement the proposal with state-of-the-art techniques. This work opens the door for quantum sensing and metrology with chiral molecules.
\end{abstract}

\maketitle

{\it Introduction---}
Controlling the quantum dynamics of molecules enables exploiting their mechanical degrees of freedom for metrology \cite{patra2020,hornberger2012, benoit2010}, for quantum information processing \cite{DeMille2002,tesch2002,rabl2006,yu2019,albert2020}, and for testing the quantum superposition principle \cite{arndt2014a}. State-of-the-art experiments range from cooling diatomic molecules into the deep ro-vibrational quantum regime \cite{staanum2010,schneider2010,lien2014, barry2014,truppe2017,anderegg2018,ding2020, wolf2016,chou2017, sinhal2020}, to coherently controlling the rotational and vibrational dynamics of molecules \cite{brinks2010,koch2019}, to centre-of-mass interference of massive molecules \cite{Fein2019a}. Simultaneously addressing the internal and external quantum dynamics of large molecules is complicated due to the vast number of interacting degrees of freedom. Here, we show how single chiral molecules can be brought into a quantum superposition of oppositely-handed configurations.

\begin{figure}[b!]
\centering
\includegraphics[width=8.75cm]{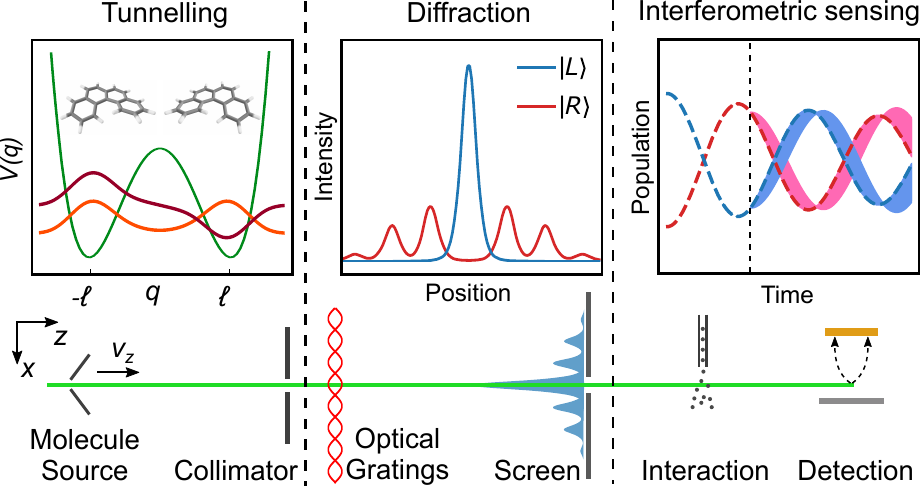}
\caption{Proposed setup to generate quantum superpositions of enantiomers and exploit their tunnelling dynamics for interferometric sensing. A racemic beam of chiral molecules is diffracted from optical gratings and filtered at the detection screen. Left inset: During the transit, the molecules continuously tunnel between left- and right-handed molecular configurations. Middle inset: Adjusting the grating phases prepares superpositions of enantiomer states. Right inset: The resulting tunnelling dynamics can be observed and exploited in subsequent beam experiments.}
\label{fig:doublewell}
\end{figure}

Understanding the properties of chiral molecules and their interaction with environments is an active interdisciplinary field of research \cite{quack2008,yukio2013}. In the absence of parity violating interactions \cite{quack1989}, their vibrational groundstate is a quantum superposition of two oppositely handed configurations, coherently tunnelling between the two mirror-images \cite{Mezey2012Book}, see Fig.\,\ref{fig:doublewell}. Preparing a molecular beam in a quantum superposition of enantiomers and observing their coherent tunnelling dynamics will open the door for interferometric measurements of handedness-dependent interactions, for enantiomeric state manipulation schemes \cite{cina1993}, for observing the tunnelling shifts due to parity-violating weak interactions \cite{harris1978,berger2001,quack2002,macdermott2004,gaul2020}, and for observing the environment-induced decay of tunnelling, as proposed to resolve Hund's paradox \cite{Hund1927,trost2009,Mezey2012Book,ghahramani2013}.

Matter-wave interference is a versatile tool for testing the quantum superposition principle with large molecules\,\cite{nimmrichter2011b,Fein2019a} and for measuring molecular properties in the gas phase\,\cite{Eibenberger2014, Mairhofer2017,Fein2019b}. In addition, far-field matter-wave interferometry combined with spatial filtering has been proposed for sorting conformers \cite{Brand2018} and enantiomers \cite{Cameron2014a,Cameron2014b} from a racemic mixture. In this article, we achieve three important goals: First, we demonstrate that far-field diffraction from a standing-wave grating together with a handedness-dependent phase shift from an optical helicity grating can be used to prepare enantiomeric superposition states of individual thermally rotating molecules. Second, we show how interferometric sensing and the environmental decay of tunnelling can be observed. Third, we identify helix-shaped [4]-helicene derivatives as suitable candidates, exhibiting tunnelling frequencies in a wide range and featuring large optical rotations. Our work presents a promising platform for investigating chiral molecules, complementing conventional spectroscopic techniques \cite{benoit2010,quack2008}.

{\it Ro-vibrational dynamics---} To illustrate the interferometric preparation of enantiomer superpositions, we consider a freely rotating chiral molecule of mass $M$ with three orientational degrees of freedom $\Omega = (\alpha,\beta,\gamma)$, e.g. Euler angles in the $z$-$y'$-$z''$ convention. The vibrational dynamics are characterized by the coordinate $q$, parametrizing the one-dimensional tunnelling path with effective mass $\mu$ and double-well potential $V(q)$. The latter gives rise to tunnelling between oppositely handed molecular configurations localized at $q = \pm \ell$.

The vibrational motion can strongly interact with the rotations via the centrifugal coupling and the Coriolis effect \cite{bunker2006}. The former is due to the configuration-dependence of the inertia tensor ${\rm I}(\Omega,q) = {\rm R}(\Omega) {\rm I}_0(q) {\rm R}^T(\Omega)$, where ${\rm I}_0(q)$ is the body-fixed inertia tensor and ${\rm R}(\Omega)$ is the rotation matrix specifying the orientation of the molecule. The Coriolis coupling is proportional to the product of vibrational velocity $\dot{q}$ and angular velocity $\boldsymbol{\omega}$, related to the time-derivative of the orientation by $\dot{\rm R}(\Omega) = \boldsymbol{\omega}\times {\rm R}(\Omega)$. 

In total, the ro-vibrational Lagrangian is of the form
\begin{equation}\label{eq:lagrangian}
    L = \frac{1}{2} \boldsymbol{\omega} \cdot {\rm I}(\Omega,q) \boldsymbol{\omega} + \kappa \dot{q}\, {\boldsymbol \omega}\cdot {\bf n}(\Omega) + \frac{\mu}{2} \dot{q}^2 - V(q),
\end{equation}
where we defined the Coriolis coupling parameter $\kappa$ and the body-fixed axis of Coriolis coupling ${\bf n} = {\bf n}(\Omega)$. The first term describes the rotational kinetic energy and centrifugal coupling, the second term describes the Coriolis effect, and the final two terms describe the body-fixed elastic motion of the molecule. Depending on the molecule, the tunnelling mass $\mu$, as well as the Coriolis coupling parameters $\kappa$ and ${\bf n}$ can depend on $q$, see App.\,\ref{app:hamiltonian}. 

The canonical vibrational and angular momentum coordinates follow from the derivatives of \eqref{eq:lagrangian} with respect to $\dot{q}$ and $\boldsymbol{\omega}$, respectively, as $p = \mu \dot{q} + \kappa \,\boldsymbol{\omega}\cdot {\bf n}$ and ${\bf J} = {\rm I}(\Omega,q) \boldsymbol{\omega} + \kappa \dot{q} {\bf n}$. Thus the mechanical deformation contributes to the total angular momentum along ${\bf n}$ and mechanical rotations around ${\bf n}$ contribute to the elastic momentum. A straight-forward calculation yields the Hamilton function
\begin{align}\label{eq:hamil}
    H = & \frac{1}{2} \left ( {\bf J} - \frac{\kappa }{\mu} p {\bf n} \right ) \cdot \Lambda^{-1}(\Omega,q) \left ( {\bf J} - \frac{ \kappa }{\mu} p {\bf n} \right ) \nonumber \\
    & + \frac{p^2}{2 \mu} +V(q),
\end{align}
with the effective tensor of inertia $\Lambda(\Omega,q) = {\rm I}(\Omega,q) - \kappa^2 {\bf n}\otimes {\bf n}/\mu$. The first term on the right-hand side describes for the rotational motion, including its centrifugal and Coriolis coupling to the vibrational dynamics

{\it Enantiomer tunnelling---} The classical Hamiltonian \eqref{eq:hamil} can be quantized by promoting all phase space variables to operators, which are denoted by sans-serif characters. The two lowest lying vibration states in the double-well potential $V(q)$ have tunnelling splitting $2\hbar \Delta$. As shown below, interferometric preparation works best if $\Delta/2 \pi$ is on the Hz to kHz level, while the spacing to the next higher vibration states is approximately THz -- a situation realistically achievable with [4]-helicene derivatives. Finally, typical rotation rates are on the order of $\omega_{\rm rot}/2\pi \simeq \, {\rm GHz}$.

This large frequency spacing between the two lowest and all higher vibration states constrains the vibrational dynamics to the left- and right-handed enantiomeric states $|L\rangle$ and $|R\rangle$, whose even and odd superpositions are the vibrational ground- and first excited states, see Fig.\,\ref{fig:doublewell}. There are thus three distinct contributions from the rotational Hamiltonian in the first line of \eqref{eq:hamil}: (i) The centrifugal coupling is approximately diagonal in the enantiomer basis because the tunnelling contribution is suppressed by $|\langle L|R\rangle| \ll 1$. (ii) The Coriolis coupling is negligible in comparison to the centrifugal coupling since the molecular rotation rate clearly exceeds the tunnelling splitting, $\Delta/\omega_{\rm rot} \ll 1$. (iii) The purely elastic contribution effectively modifies the tunnelling splitting from $\Delta$ to $\Delta_*$.

We explicitly derive the ro-vibrational Hamiltonian \eqref{eq:hamil} and $\Delta_*$ by performing the two-level approximation for a half- and full helix in Apps.\,\ref{app:helix} and \ref{app:twolevel}. The helix model can be used to estimate the influence of ro-vibrational couplings for helix-shaped $[n]$-helicene molecules.

\begin{figure*}[t]
\centering
\includegraphics[width=18cm]{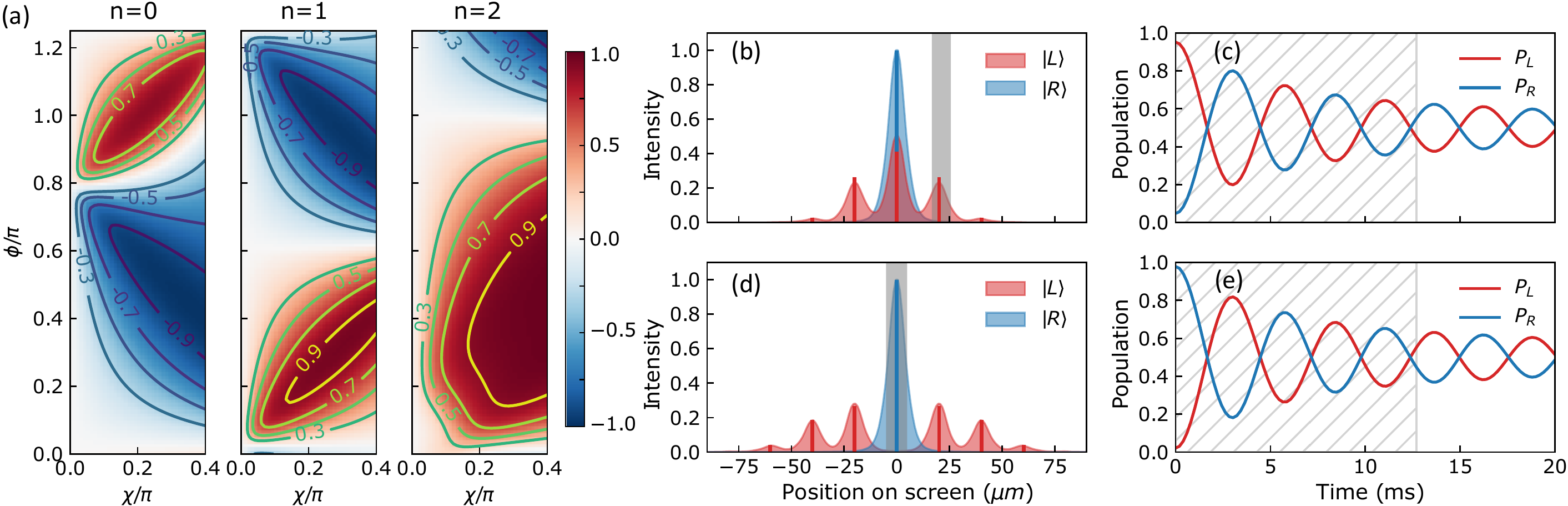}
\caption{Enantio-selectivity for [4]-helicene derivatives: (a) Numerically computed difference of left- and right-handed molecule populations after spatially filtering the zeroth (left), first (middle) and second (right) diffraction orders as a function of the grating phases $\phi$ and $\chi$ [prefactors of Eqs.\,\eqref{eq:gratephase} and \eqref{eq:helphase}]. Blue and red refer to an excess of right- and left-handed enantiomers, respectively, indicating that high enantio-selectivity can be achieved in all diffraction orders. (b) Diffraction pattern for $(\phi, \chi)=(0.21\pi, 0.19\pi)$. Red and blue shaded areas depict the numerically computed diffraction pattern for the $|L\rangle$ and $|R\rangle$ states. The solid lines denote the position and amplitude of Eq.\,\eqref{eq:coeff}. (c) Evolution of the enantiomer populations $P_{L/R}(t)$ from the grating to the screen (hatched background) and after spatial filtering (blank background). The evolved states are selected by filtering the first diffraction order, as indicated by the grey-shaded area in (b). The influence of rotations is calculated with the full-helix model at a rotational temperature of $4$\,K. Plots (d) and (e) show the same as (b) and (c) but for  selecting the zeroth diffraction order at $(\phi, \chi)=(0.39\pi, 0.38\pi)$.}
\label{fig:diffractionpattern}
\end{figure*}

In summary, the Hamiltonian in the vibrational two-level approximation reads
\begin{align} \label{eq:twolevelham}
    {\sf H} \simeq & -\hbar \Delta_* \left (|L\rangle \langle R| + {\rm h.c.} \right ) \nonumber\\
    & + {\sf H}_L \otimes |L\rangle \langle L| + {\sf H}_R \otimes |R\rangle \langle R|
\end{align}
Here, the free rotational Hamiltonians ${\sf H}_{L} = {\boldsymbol{ \sf J}} \cdot {\sf \Lambda}_{L}^{-1} {\boldsymbol{ \sf J}}/2$, and likewise for $R$, describes the motion of an asymmetric rotor with inertia tensor $\Lambda_{L/R} =\Lambda(\Omega,\pm\ell)$. These tensors have identical eigenvalues but distinct eigenvectors. They are thus related to each other by a rotation of the body-fixed frame.

The Hamiltonians ${\sf H}_{L/R}$ share the same eigen-energies $E_{\eta}$, where $\eta = (jn)$ labels the $n$-th rotation eigenstate with total angular momentum $j$. The free rotational Hamiltonian is isotropic and thus its eigenstates and eigenenergies are independent of the space-fixed angular momentum quantum number $m$. However, the eigenstates $|\eta; {L/R}\rangle$ are related by a rotation of the body-fixed frame. The relation between left- and right-handed rotation states is given in App.\,\ref{app:rotstates} for the half- and full helix. The rotational eigenstates $|\eta;L/R\rangle$ at fixed molecular configuration must not be confused with the configuration states $|L/R\rangle$, which cannot be interconverted by a rotation. 

The rotations can be eliminated adiabatically by performing a rotating-wave approximation since the rotation rates significantly exceed the tunnelling splitting. The resulting Hamiltonian describes combined ro-vibrational tunnelling between states of identical rotational energy but opposite handedness,
\begin{equation} \label{eq:rovibham}
   {\sf H} \simeq - \hbar \sum_{\eta} \Delta_{\eta} e^{i\gamma_{\eta}} |\eta L \rangle \langle \eta R|  + {\rm h.c.},
\end{equation}
where $|\eta L\rangle = |jn;L\rangle |L\rangle$ and likewise for $R$. Thus each rotation state $\eta$ yields a distinct tunnelling frequency, which depends on the overlap of rotation states, $\Delta_{\eta} e^{i \gamma_{\eta}} = \Delta_*  \langle \eta;L | \eta;R\rangle$. Physically, this describes that the enantiomeric states and the molecular rotations become entangled by the free dynamics with the Hamiltonian \eqref{eq:twolevelham}.

The eigenstates of \eqref{eq:rovibham} with energies $\mp \hbar \Delta_{\eta}$ are given by $|\psi_{\eta \sigma}\rangle = (|\eta L\rangle -\sigma e^{-i\gamma_\eta} |\eta R\rangle)/\sqrt{2}$ with $\sigma = \mp$. The resulting ro-vibrational tunnelling dynamics of the pure states $|\eta L\rangle$ and $|\eta R\rangle$ follow as
\begin{subequations} \label{eq:timeevolv}
\begin{align}
    |\Psi_{\eta L} \rangle = & \cos(\Delta_{\eta} t) |\eta L\rangle + i e^{-i \gamma_{\eta}} \sin(\Delta_{\eta} t) |\eta R\rangle , \\
    |\Psi_{\eta R}\rangle = & \cos(\Delta_{\eta} t) |\eta R\rangle + i e^{i \gamma_{\eta}} \sin(\Delta_{\eta} t) |\eta L\rangle.
\end{align}
\end{subequations}
In what follows, we will show how these tunnelling dynamics can be combined with centre-of-mass molecule interferometry to prepare enantiomer superposition states.

{\it Matter-wave diffraction---} A beam of molecules freely propagates with constant forward velocity $v_z$ from the source to the optical grating, traverses the light field, and continues freely to the interference screen at distance $d$ further downstream, see Fig.~\ref{fig:doublewell}. At the source, the molecular centre-of-mass state is uncorrelated with the internal degrees of freedom and described by Gaussian smeared point sources \cite{Brand2018}. The ro-vibrational degrees of freedom are in a thermal state of temperature of 4\,K and with the Hamiltonian \eqref{eq:rovibham}. The centre-of-mass state disperses during the passage to the grating, while the internal state does not change. 

The centre-of-mass and ro-vibrational states become entangled through diffraction at two optical gratings. The gratings are each generated by a single laser pulse, which is split into two beams and recombined: the first pulse generates a standing-wave grating, imprinting the spatial phase modulation $\phi_\eta^{L/R}(x)$ on the centre-of-mass state. The second pulse realises a helicity grating \cite{Cameron2014a,Cameron2014b}, introducing the enantiomer-specific phase modulation $\chi_\eta^{L/R}(x)$. The combined action of both pulses can be described by a grating transformation operator, which is diagonal in the transverse centre-of-mass position as well as in the ro-vibrational states,
\begin{equation} \label{eq:grattrafo}
    {\sf t} = \sum_{\eta} \left ( t_\eta^L({\sf x}) \otimes |\eta L\rangle\langle \eta L| + t_{\eta}^R({\sf x}) \otimes |\eta R\rangle\langle \eta R|\right ).
\end{equation}

The grating transformation operators can be calculated in the eikonal approximation \cite{stickler2015,Brand2018,Cameron2014a}. For molecules with negligible absorption this yields $t^{L}_\eta(x) = \exp [i\phi_\eta^{L}(x) + i \chi^{L}_\eta(x)]$, where $x$ must lie within the grating of width $w$, and likewise for $R$. For transverse positions $x$ outside the grating, $t_\eta^L(x) = 0$, because the molecular beam is blocked by a collimation aperture, see Fig.\,\ref{fig:doublewell}. The phase due to the pulsed standing-wave grating, generated by overlapping two co-propagating laser beams at angle $2\theta$ with wavelength $\lambda$ and the same linear polarization, is given by
\begin{equation}\label{eq:gratephase}
\phi_\eta^{L}(x)=\sqrt{\frac{\pi}{2}} \dfrac{ \alpha_\eta^L \tau E^2 }{\hbar}\left [1+\cos\left (4 \pi \sin\theta \frac{x}{\lambda}\right )\right] .
\end{equation}
Here, $E$ is the amplitude of a Gaussian pulse of duration $\tau$, $E(t) = E \exp (- t^2/\tau^2)$ and $\alpha_\eta^L$ is the expectation value of the polarisability tensor of state $|\eta;L\rangle$ in direction of the laser polarisation. For large and cold molecules, the influence of the polarizability anisotropy on interference is typically small \cite{stickler2015}, so that we can replace $\alpha_\eta^L$ by the rotationally averaged polarizability $\alpha$, which is independent of the molecule handedness \cite{barron2009}.

The helicity grating is generated by superposing two laser beams at angle $2\theta$ but with orthogonal linear polarizations and electric field strength $E_{\rm hel}$. This yields the eikonal phase \cite{Cameron2014a}
\begin{equation}\label{eq:helphase}
\chi_\eta^{L}(x)=\sqrt{\frac{\pi}{2}}\dfrac{ G_\eta^L  \tau E_{\rm hel}^2\cos^2{\theta}}{\hbar c}\sin\left (4 \pi \sin \theta\frac{x}{\lambda} + \vartheta \right ),
\end{equation}
and similar for $R$. The transverse spatial offset between the standing-wave and the helicity gratings is quantified by $\vartheta$. Again, we approximate the $|\eta;L\rangle$-expectation values $G_\eta^L$ of the electric-magnetic dipole polarizability pseudotensor \cite{barron2009} by its rotational average $G_L$. Since it depends on the molecular handedness, $G_L = - G_R$, the helicity grating imprints phases of opposite sign on left- and right-handed states. Thus, the centre-of-mass and ro-vibrational states become entangled at the gratings, enabling enantiomer selection further downstream.

The diffraction pattern follows from freely propagating the total ro-vibrational and centre-of-mass state from the grating to the screen and tracing out the ro-vibrational degrees of freedom. Denoting the combined ro-vibrational and centre-of-mass unitary time evolution operators from the source to the grating by ${\sf U}_1$ and from the grating to the screen by ${\sf U}_2$, the final diffraction pattern for $L$-enantiomers and the centre-of-mass source state $|x_0\rangle$ reads
\begin{equation}\label{eq:diff}
    S_L(x,x_0) = \frac{1}{Z} \sum_{\eta \sigma}e^{-\beta \epsilon_{\eta \sigma}} \left | \left \langle \eta L \right \vert \langle x \vert {\sf U}_2 {\sf t} {\sf U}_1 \vert \psi_{\eta \sigma} \rangle \vert x_0 \rangle \right |^2,
\end{equation}
where $Z$ is the ro-vibrational partition function and similar for $R$. The trace over the ro-vibration states is weighted with the Boltzmann factor of energy $\epsilon_{\eta \sigma} = E_\eta + \sigma \hbar \Delta_\eta$ and temperature $1/k_{\rm B} \beta$. Finally, one must average \eqref{eq:diff} over an Gaussian ensemble of centre-of-mass point sources.

{\it Enantiomer superpositions---} The resulting enantiomer state at the screen can be approximated analytically in the far field. For this purpose, we consider that the centre-of-mass state before the grating is given by the transverse plane wave $|p_x = 0\rangle$, which then propagates with constant forward velocity $v_z$ towards the screen at distance $d$. Starting out in the ro-vibrational state $|\psi_{\eta \sigma} \rangle$, the total state at the screen is $(|\Psi_{\eta L}\rangle |\Phi_L \rangle - \sigma e^{- i \gamma_\eta} |\Psi_{\eta R}\rangle |\Phi_R \rangle)/\sqrt{2}$. Here, the ro-vibrational states are given by Eq.~\eqref{eq:timeevolv} with $t = d/v_z$ the time from the grating to the screen. Similarly, the centre-of-mass states $|\Phi_{L/R}\rangle =  {\sf U}^{(\rm cm)}_2 t^{L/R}({\sf x}) |p_x = 0\rangle$ have been diffracted with the $L$- and $R$-grating transformations \eqref{eq:grattrafo}, respectively, where ${\sf U}_2^{(\rm cm)}$ describes the free centre-of-mass time evolution from the grating to the screen.

In the far-field regime, $m v_z w^2/8\hbar d \ll 1$, individual diffraction orders of the spatial wave functions $|\Phi_{L/R}\rangle$ are well separated by $4\pi \hbar d \sin \theta /M v_z \lambda$. Tuning the relative grating offset to $\vartheta = \pi/2$, the amplitude of the $n$-th diffraction order is given by
\begin{equation}\label{eq:coeff}
    c^{L}_n \propto i^n J_n \left [ \sqrt{\frac{\pi}{2}}\frac{\alpha\tau E^2}{\hbar } \left ( 1  + \frac{G_L}{c \alpha}\frac{E_{\rm hel}^2 }{E^2} \cos^2\theta\right )\right ] ,
\end{equation}
and likewise for $R$, where $J_n(\cdot )$ denotes the $n$-th order Bessel function and we dropped the normalization, see App.~\ref{app:ffdiff}. Spatially filtering the $n$-th diffraction order from the full interference pattern would thus prepare the ro-vibrational superposition state $c_n^L |\Psi_{\eta L} \rangle - \sigma e^{-i \gamma_\eta} c_n^R |\Psi_{\eta R}\rangle$ from the pure initial state $|\psi_{\eta \sigma}\rangle$.

The amplitudes \eqref{eq:coeff} depend on the molecular handedness through the sign of the electric-magnetic dipole coupling $G_L = - G_R$. Thus, the relative weight of the left- and right-state in the filtered wave packet can be set via the ratio of laser intensities $E_{\rm hel}^2/E^2$ as well as the distance $d$ between the grating and the screen. For instance, tuning the intensity ratio such that $c_n^R \gg c_n^L$ prepares the state
\begin{equation} \label{eq:rho}
    \rho \simeq \sum_{\eta} p_\eta |\Psi_{\eta R}\rangle \langle \Psi_{\eta R} |,
\end{equation}
with statistical weights $p_\eta = \sum_\sigma \exp (-\beta \varepsilon_{\eta \sigma})/Z \simeq \exp( - \beta E_\eta)/Z$. The spatial filtering thus post-selects states of a definite handedness at the grating. The resulting interference patterns for a set of experimentally feasible parameters are illustrated in Fig.~\ref{fig:diffractionpattern}(b) and (d), showing remarkable agreement between the exact expression \eqref{eq:diff} and the analytic approximation \eqref{eq:coeff}. Note that area of the interferograms from left- and right-handed molecules are equal, while they distribute differently across the diffraction orders. The enantiomer-excess in Fig.\,\ref{fig:diffractionpattern}(a) demonstrates that matter-wave diffraction can be used for creating enantiomer superpositions.

The enantiomeric composition of the molecular beam at a variable distance behind the screen can be measured with state-of-the-art photoionization and velocity-map imaging techniques \cite{Kastner2016}. In the absence of decoherence, the resulting probability of observing the molecule in its $L$ configuration at the distance $v_z t$ behind the interference grating is given by $P_R(t) = \sum_\eta p_\eta \cos^2 (\Delta_\eta t)$. It decays with time and approaches $1/2$ on a timescale determined by the ro-vibrational coupling. This is calculated with the full-helix model and shown in Fig.~\ref{fig:diffractionpattern}(c,e).

{\it Interferometric sensing---} Having prepared molecules in a superposition of enantiomer states, the setup enables interferometric sensing and metrology of handedness-dependent external forces. For instance, post-selecting the $R$-enantiomer at the detection screen, the state at distance $v_z t$ after the grating is given by Eq.\,\eqref{eq:rho}. If the molecules interact with the potential ${\sf V}_{\rm int}({\boldsymbol{\sf r}}_\bot, {\sf z}) = V_L({\boldsymbol{\sf r}}_\bot, {\sf z}) |L\rangle\langle  L| + V_R({\boldsymbol{\sf r}}_\bot, {\sf z})|R\rangle\langle R|$ for a short period $\tau_{\rm int} \ll 1/\Delta_\eta$, where $z$ denotes the centre-of-mass position of the molecule along the flight direction and ${\bf r}_\bot$ is its transverse position, the $L$- and $R$-enantiomers accumulate the relative phase
\begin{equation}
    \varphi \simeq \frac{1}{\hbar} \int_{-\infty}^\infty dt'\, \left [V_R({\bf r}_\bot, v_z t') - V_L({\bf r}_\bot, v_z t')\right ].
\end{equation}
The probability of measuring the $R$ configuration after another free time evolution over the distance $v_z t$ is then modified by the presence of the potential according to $\overline{P}_R(t,t) = 1 - \cos^2(\varphi/2) [ 1 - P_R(2t)]$, as detailed in App.\,\ref{app:sensing}, where $P_R(2t)$ is the probability without the phase shift. The modification becomes observable for phase differences on the order of $\pi/2$, amounting to energy differences of $10^{-27}$\,Joule and force differences of $10^{-23}$\,N, for $\tau_{\rm int} \simeq 1\,\mu$s and $v_z \simeq 300$\,m/s.

Interferometric sensing with a beam of chiral molecules can be used to measure enantiomer-dependent forces due to nearby surfaces, other chiral molecules, or optical fields with unprecedented accuracy. For instance, letting the molecules fly briefly at distance $d$ to a chiral surface induces the phase difference $\varphi \simeq \sum_i R_{0i}[1/3+\ln{(\omega_{0i}c/d})]\tau_{\rm int} /6 \pi^2 \hbar \varepsilon_0  d^3 $ \cite{barcellona2016}, where $R_{0i}$ and $\omega_{0i}$  are the rotatory strengths and frequencies of the electronic transitions. For a separation of $d \simeq 100\,$nm and considering the dominant electronic transitions of [4]-helicene \cite{furche2000} this yields $\varphi \simeq \pi/2$. In a similar fashion, this setup can be exploited for interferometric metrology of chiral molecules. For instance, the electric-magnetic dipole cross-polarizability tensor can be measured by illuminating the molecular beam with pulses of chiral light \cite{canaguier2013}, which leads to differential interaction with the $L/R$ enantiomers.

In addition, imprinting the relative phase $\varphi = \pi$ performs motional decoupling, reversing rotational phase averaging of the tunnelling dynamics.

\begin{figure}
    \centering
    \includegraphics[width = 0.35\textwidth]{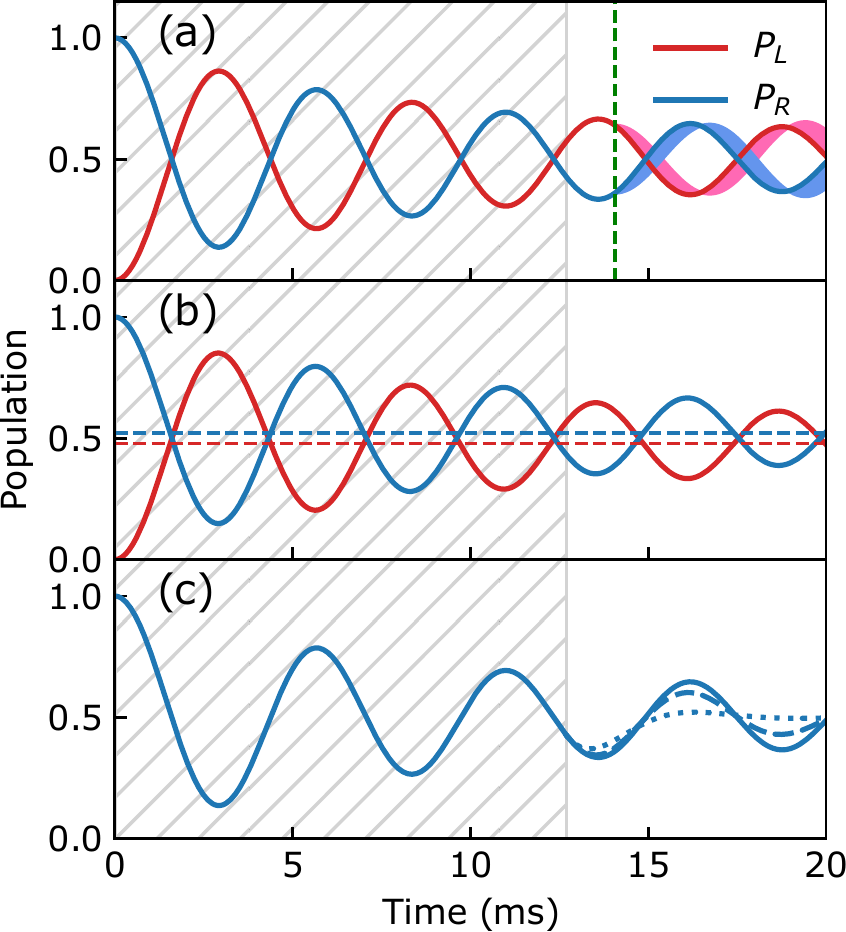}
    \caption{(a) Interferometric sensing of enantio-selective forces: Imprinting relative phases $\varphi \leq \pi/2$ after the screen (vertical dashed line) strongly affects the subsequent tunnelling dynamics, as indicated by the shaded region. (b) Measuring parity violation: The presence of parity violating interactions shifts the mean values of the probabilities for observing left- and right-handed enantiomers, as indicated by the dashed lines. (c) Environmental super-selection of handed enantiomers: The environment-induced decay of enantiomer tunnelling can be probed in a dedicated collision chamber starting right after the screen. The solid line shows the free tunnelling dynamics (evacuated  collision chamber), while dashed and dotted show the influence of $N_2$ molecules at pressures $1\times10^{-6}$\,mbar and $5\times10^{-6}$\,mbar, respectively.}
    \label{fig:deco}
\end{figure}

{\it Sensing of parity violation---} Detecting parity-violating weak interaction through enantiomer tunnelling was proposed in Ref.\,\cite{harris1978}. Here, we calculate how the modified tunnelling dynamics influence the measurement signal. The parity violating contribution to the molecular vibration state adds
\begin{equation}
    {\sf H}_{\rm pv} = \hbar \omega_{\rm pv} \sum_{\eta} \left ( |\eta L\rangle \langle \eta L| - |\eta R\rangle \langle \eta R| \right ),
\end{equation}
to the Hamiltonian \eqref{eq:hamil} \cite{harris1978,macdermott2004}.
This modifies the tunnelling frequency, $\Omega_\eta^2 = \omega_{\rm pv}^2 + \Delta_\eta^2$, and introduces a phase offset, replacing Eqs.\,\eqref{eq:timeevolv} by
\begin{subequations}
\begin{align}
    |\Psi_{\eta L}'\rangle = & \left [ \cos(\Omega_\eta t) - i \frac{\omega_{\rm pv}}{\Omega_\eta} \sin (\Omega_\eta t)
      \right ] |\eta L\rangle \nonumber \\
      & + i \frac{\Delta_\eta}{\Omega_\eta} \sin(\Omega_\eta t ) e^{-i \gamma_\eta} |\eta R\rangle
\end{align}
\begin{align}
    |\Psi_{\eta R}'\rangle = & \left [ \cos(\Omega_\eta t) + i \frac{\omega_{\rm pv}}{\Omega_\eta}\sin (\Omega_\eta t)       \right ] |\eta R\rangle \nonumber \\
      & + i  \frac{\Delta_\eta}{\Omega_\eta} \sin(\Omega_\eta t )   e^{i \gamma_\eta}|\eta L\rangle
\end{align}
\end{subequations}
The resulting probability $P_R'(t)$, after post-selecting the $R$-enantiomer, oscillates with reduced amplitude around the shifted mean of $1/2 + \sum_\eta p_\eta \omega_{\rm pv}^2/2\Omega_\eta^2$, see Fig.\,\ref{fig:deco}(b).

In naturally occurring molecules, parity violation due to the electro-weak interaction slightly lifts the degeneracy of the $L$- and $R$-enantiomers on the sub-mHz level \cite{harris1978,quack1986,quack1989}. While such small splittings are not observable with this setup, molecules containing heavy elements can reach splittings of up to $\sim10$\,Hz \cite{berger2019}. Assuming $\omega_{\rm pv}/2\pi = 10\,$Hz and $\Omega_\eta/2\pi \simeq 100$\,Hz, the means of $P_{L/R}'(t)$ are separated by $\simeq1\,\%$, as shown in Fig.\,\ref{fig:deco}(b), which can be resolved with state-of-the-art photo-electron circular dichroism measurements \cite{Kastner2016}.

{\it Environmental decay of tunnelling---} In addition to interferometric sensing, the proposed setup also enables observing the tunnelling decay under environmental interactions, which has been proposed to resolve Hund's paradox \cite{Hund1927,Mezey2012Book}. In the presence of a background gas, the free motion of the molecules is interrupted by random collisions with individual gas particles. These scattering processes exert linear and angular momentum kicks to the molecules, thereby localizing their position and orientation \cite{stickler2016b,papendell2017}. The resulting decay of rotational coherences implies decoherence of the enantiomer superpositions, since the rotational and elastic degrees of freedom are entangled, see Eq.\,\eqref{eq:timeevolv}. The decoherence rate of enantiomer superpositions can thus be estimated by the total rate of collisions with gas particles of mass $m$ and density $n_g$, $\Gamma_{LR} \simeq n_g \langle p  \sigma_{\rm tot}(p) \rangle/m$. Here, the average is taken over the Maxwell-Boltzmann distribution of gas momenta $p$ at temperature $T$ and $\sigma_{\rm tot}(p)$ is the total scattering cross section of van der Waals scattering.

The total scattering cross section can be calculated in the eikonal approximation via the optical theorem for highly-retarded van der Waals scattering with the rotationally averaged potential $V(r) = -C/r^7$, where the van der Waals constant depends on the gas polarizability $\alpha_g$ as $C = 23 \hbar c \alpha_{\rm g} \alpha/64 \varepsilon_0^2\pi^3$ \cite{buhmann1,Craig1998Book}. A straight-forward calculation based on Ref.\, \cite{stickler2016b} yields
\begin{equation} \label{eq:decorate}
    \Gamma_{LR} \simeq 5 n_g \Gamma\left (\frac{2}{3}\right )\Gamma \left ( \frac{5}{6} \right ) \sqrt{\frac{2 \pi k_{\rm B}T}{3 m}} \left ( \frac{16 \sqrt{m} C}{15 \hbar \sqrt{2 k_{\rm B} T}} \right )^{1/3},
\end{equation}
adding an exponential decay to the chiral tunnelling dynamics of the reduced state \eqref{eq:rho}, see App.\,\ref{app:deco}.

The resulting tunnelling dynamics for various pressures of N${}_2$ background gas are shown in Fig.\,\ref{fig:deco}, where we assumed that the region after spatial filtering is homogeneously filled with the gas. This demonstrates that the environmental suppression of chiral tunnelling is observable under realistic experimental conditions.

{\it Identifying suitable molecules---} Interferometric preparation of enantiomer superposition requires (i) strong chiro-optical response for achieving sufficiently large phase shifts at the interference grating and (ii) groundstate splittings in the range of $10$\,Hz to $50$\,kHz, where tunnelling dynamics can be realistically observed during the transit from the screen to the detector. Tunnelling splitting calculations in this regime are computationally expensive because of the large size of the molecules and the required numerical accuracy. We perform extensive calculations with density functional theory and the instanton approach (see App.\,\ref{app:methods}) to identify [4]-helicene derivatives as a promising class of molecules.

The optical properties of [4]-helicene render it suitable for interferometric enantio-separation, given that $\alpha=3.8\times10^{-39}\,\mathrm{C m^2/V}$ and $G_{L/R}=\pm7.0\times10^{-36}\,\mathrm{m A^2 s^3/kg}$ at $\lambda = 1$\,$\mu$m. The high vibrational frequency $\sim 2~\mathrm{THz}$ justifies the two-level approximation at 4\,K. The electronic barrier for inversion is $\sim 50~\mathrm{THz}$, rendering the instanton approach \cite{richardson2011,sahu2020} appropriate for calculating the tunneling splitting. These calculations yield $\Delta/2\pi \sim 1\,$pHz for [4]-helicene, which is too small for the proposed experiment. The tunnelling splitting of helicenes can be significantly increased by chemical substitutions, which only affects mildly their optical properties \cite{janke1996,ravat2017}. Calculations for 1H-naphtho[2,1-g]indole (naphthindole) yield $\Delta/2\pi \sim 1 - 200~\mathrm{MHz}$, while $\alpha$ and $G_{L/R}$ remain on the same order of magnitude. These two compounds bracket the desired frequency range and thus provide an excellent starting point for future investigations. We suggest replacing the side groups to slightly change the tunnelling barrier, while using additional substituents or isotopes in the periphery of the molecule to alter the effective tunneling mass.

{\it Proposed experiment---} The required molecular beam can be generated in a seeded supersonic jet \cite{wall2016} or from a buffer gas cell \cite{hutzler2012}, yielding ro-vibration temperatures of a few Kelvin. The laser field strengths required for diffraction can be achieved by a pulsed Nd:YAG laser operating at $1064$\,nm, exhibiting a pulse energy of 1\,J, a pulse duration of 10\,ns, and a beam diameter of 0.5\,mm. The laser wavelength lies in the transparent region of [4]-helicene derivatives to minimize photon absorption. In addition, molecules, which absorbed a photon during the grating transit, can be filtered out at the screen because of the ensuing momentum kick \cite{Brand2018}. The beam crossing angle $\theta=0.5$ implies a grating period of $1.1\,\mu$m, yielding a rather long total interferometer length of $7$\,m at $v_z\simeq315$\,m/s. The resulting interference pattern and enantiomer tunnelling dynamics are illustrated in Figs.\,\ref{fig:diffractionpattern} and \ref{fig:deco}, see App.\,\ref{app:experiment} for further details. The enantiomeric composition of the final beam can be measured with high resolution using photo-electron circular dichroism measurement \cite{Kastner2016} or Coulomb explosion \cite{pitzer2013}. In both detection methods, the molecules can be ionized using a laser spot of approximate size $10\,\mu$m, which when extended over the vertical direction of the molecular beam implies a longitudinal velocity spread of $0.4$\,m/s. Averaging the tunnelling signal over this velocity spread shows that enantiomer tunnelling is observable for molecules with tunnelling frequencies smaller than $50$\,kHz, thus complementing existing spectroscopic techniques \cite{benoit2010,quack2008,patterson2013,domingos2017}. This interferometric scheme can reach enantiomeric purities beyond 90\% at relatively small helicity-grating phase shifts on the order of $\pi/10$. 

{\it Conclusion---} In conclusion, we demonstrated that far-field matter-wave diffraction can be used to prepare enantiomer superpositions with feasible experimental parameters. This will open the door to the observation of enantiomer tunnelling and its exploitation for sensing and metrology, including the determination of chiral Casimir-Polder forces and potentially parity-violating effects inside the molecule. We identify [4]-helicene derivatives as suitable molecular candidates. Extended ab-initio calculations will help to further narrow this down to a specific compound. Future work could investigate how interferometric preparation of enantiomer superpositions can be achieved in near-field interferometers \cite{hornberger2012} or trapped molecules, potentially alleviating the requirements on the molecular properties. In addition, the presented setup can in principle also be used to observe and manipulate the tunnelling dynamics between distinct conformations of achiral molecules.

{\it Acknowledgements---} We thank Klaus Hornberger, Melanie Schnell, and Kilian Singer for stimulating discussions and feedback on the manuscript. This work is supported by Deutsche Forschungsgemeinschaft (DFG, German Research Foundation) through CRC 1319.

\begin{widetext}
\appendix

\section{Derivation of the ro-vibrational Hamiltonian}\label{app:hamiltonian}

We will show how the ro-vibrational Lagrangian
\begin{equation}\label{eq:lag}
    L = \frac{1}{2} \sum_{n = 1}^N m_n \dot{\bf r}_n^2 - V({\bf r}_1,\ldots,{\bf r}_N),
\end{equation}
describing the dynamics of $N$ atoms with masses $m_n$ and centre-of-mass positions ${\bf r}_n$ interacting via the molecular potential $V({\bf r}_1,\ldots,{\bf r}_N)$, can be simplified in the limit that only a single vibrational degree of freedom $q$ matters \cite{bunker2006}. In this case, the atomic positions are fully determined by the value of $q$ and by the molecule orientation $\Omega$, 
\begin{equation}\label{eq:rn}
{\bf r}_n (\Omega,q) = {\rm R}(\Omega) {\bf r}_n^{(0)}(q),    
\end{equation}
where ${\rm R}(\Omega)$ is the matrix rotating between body- and space-fixed frames and ${\bf r}_n^{(0)}(q)$ are the atom positions in the body-fixed frame. The resulting velocities are related to the angular velocity vector $\boldsymbol{\omega}$ and to $\dot{q}$ by
\begin{equation}\label{eq:rnd}
    \dot{\bf r}_n(\Omega,q) = \boldsymbol{\omega}\times {\bf r}_n(\Omega,q) + \dot{q} {\rm R}(\Omega) [\partial_q {\bf r}_n^{(0)}(q)].
\end{equation}

Assuming that there are no external forces and torques, $V({\bf r}_1,\ldots,{\bf r}_N) = V(q)$, and plugging Eqs.\,\eqref{eq:rn} and \eqref{eq:rnd} into the Lagrangian \eqref{eq:lag} yields the Lagrangian \eqref{eq:lagrangian} with the effective mass
\begin{subequations}
\begin{equation}
    \mu(q) = \sum_{n = 1}^N m_n [\partial_q {\bf r}_n^{(0)}(q)]^2,
\end{equation}
the inertia tensor
\begin{equation}
    {\rm I}(\Omega,q) = {\rm R}(\Omega)\left [\sum_{n = 1}^N m_n \left ( [{\bf r}_n^{(0)}(q)]^2 \mathds{1} - {\bf r}^{(0)}_n(q) \otimes {\bf r}^{(0)}_n(q) \right ) \right ] {\rm R}^T(\Omega),
\end{equation}
and the Coriolis-coupling constants
\begin{equation}
    \kappa(q) = \left | \sum_{n = 1}^N m_n {\bf r}_n^{(0)}(q) \times [\partial_q {\bf r}_n^{(0)}(q)] \right |,
\end{equation}
\begin{equation}
    {\bf n}(\Omega,q) = \frac{1}{\kappa(q)} {\rm R}(\Omega) \left [ \sum_{n=1}^N m_n {\bf r}_n^{(0)}(q) \times [\partial_q {\bf r}_n^{(0)}(q)] \right ].
\end{equation}
\end{subequations}
In the following, we will drop the arguments for brevity.

The Lagrangian \eqref{eq:lagrangian} yields the canonical momenta $p = \partial L/\partial \dot{q} = \mu \dot{q} + \kappa \boldsymbol{\omega}\cdot {\bf n}$ and ${\bf J} = \partial L/\partial \boldsymbol{\omega} = {\rm I}\boldsymbol{\omega} + \kappa\dot{q} {\bf n}$. Solving the resulting relations for the velocities yields
\begin{subequations}
\begin{equation} \label{eq:omega}
    \boldsymbol{\omega} = \Lambda^{-1} \left ( {\bf J} -  \frac{\kappa}{\mu} p {\bf n}\right )
\end{equation}
and
\begin{equation}
    \dot{q} = \frac{p}{\mu} - \frac{\kappa}{\mu}  \boldsymbol{\omega} \cdot {\bf n}.
\end{equation}
\end{subequations}
Here, we defined the effective inertia tensor
\begin{equation}
    \Lambda = {\rm I} - \frac{\kappa^2}{\mu} {\bf n} \otimes {\bf n}.
\end{equation}
With these relations one can carry out the Legendre transformation,
\begin{equation}
    H = \dot{q}p + \boldsymbol{\omega}\cdot{\bf J} - L = \frac{1}{2}\dot{q}p + \frac{1}{2}\boldsymbol{\omega}\cdot{\bf J},
\end{equation}
yielding the Hamiltonian
\begin{equation}\label{eq:ham1}
    H = \frac{1}{2} \left ( {\bf J} - \frac{\kappa}{\mu} p {\bf n} \right ) \cdot \Lambda^{-1} \left ( {\bf J} - \frac{\kappa}{\mu} p {\bf n} \right ) + \frac{p^2}{2 \mu} +V(q).
\end{equation}

Both ro-vibrational canonical momenta $p$ and ${\bf J}$ contain contributions due to mechanical rotations and deformations. As a consequence, the limit of no mechanical rotation is described by the condition that ${\bf J} = \kappa p{\bf n} / \mu$, see Eq.\,\eqref{eq:omega}, rather than ${\bf J} = 0$.

\begin{figure}
\includegraphics[width=0.8\linewidth]{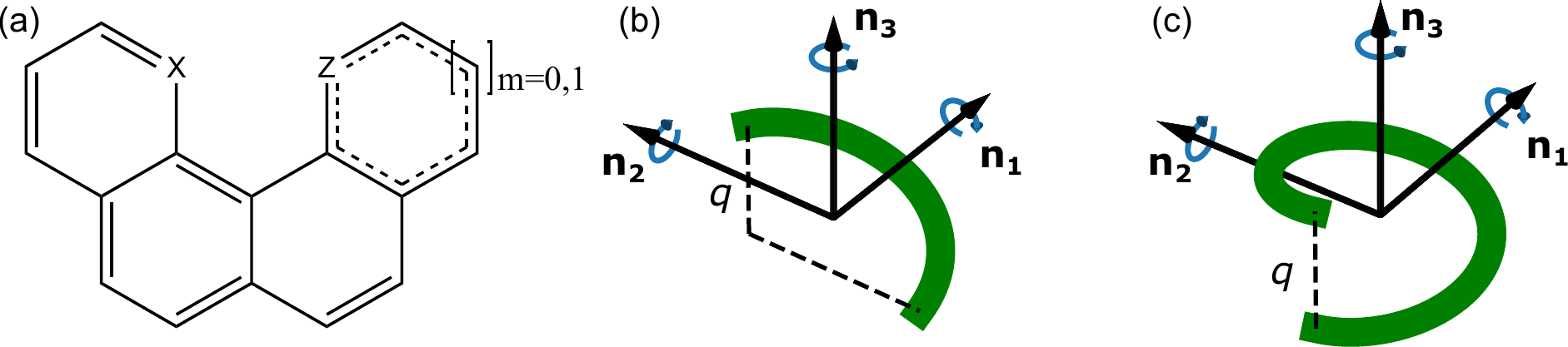}
\caption{(a) Skeletal formula of the [4]-helicene derivatives considered herein. The parent [4]-helicene is obtained when X and Z, which mark the fjord positions, correspond to C--H groups and $m=1$, so that four fused six-membered rings are obtained. For $m=0$, the ring containing Z becomes a five-membered ring, as in the naphthindole (X=C--H, Z=N, $m=0$). Variation of X, Z and $m$ has a strong impact on the barrier for stereomutation and the corresponding tunneling splittings. (b) Illustration of a half-helix body used for modelling the ro-vibrational coupling. (c) Same as in (b) but for a full helix.
\label{fig:helicenestruct}}
\end{figure}

\section{Half- and full-helix Hamiltonian}\label{app:helix}

{\it Half helix---} To quantify the effect of rotations on the tunnelling dynamics for [4]-helicene derivatives, we perform a continuum approximation, which models the latter as homogeneous helix wires of total mass $M$, radius $r$, and small opening $q \ll r$. Denoting the body-fixed helix axis by ${\bf n}_3$ and its body-fixed opening axis as ${\bf n}_1$, the  helix is in the center-of-mass frame described by $\boldsymbol{\xi}(\varphi) = r ( {\bf n}_1 \cos \varphi + {\bf n}_2 \sin \varphi ) - 2r {\bf n}_1/\pi + \varphi q {\bf n}_3 / \pi$, where the angle $\varphi \in [-\pi/2,\pi/2]$ parametrizes the helix, see Fig.\,\ref{fig:helicenestruct}. The opening $q$ determines whether the helix is left-handed ($q < 0$) or right-handed ($q>0$) handed and ${\bf n}_2 = {\bf n}_3 \times {\bf n}_1$.

The molecule's angular velocity $\boldsymbol \omega$ quantifies the rotation of the body-fixed frame $\{{\bf n}_1, {\bf n}_2, {\bf n}_3\}$ as $\dot{\bf n}_k = {\boldsymbol \omega} \times {\bf n}_k$. Together with the elastic deformations of the helix along the helix axis ${\bf n}_3$, this determines the velocity of each infinitesimal wire segment, $\dot{\boldsymbol{\xi}}(\varphi) = \varphi \dot{q} {\bf n}_3/ \pi + {\boldsymbol \omega} \times {\boldsymbol \xi} (\varphi)$. Note that the chosen parametrization of the helix implies ${\bf J}_{\rm vib} \propto Mr \dot{q} {\bf n}_1 \neq 0$, not fulfilling the Eckart equation since there is no unique vibrational equilibrium \cite{bunker2006}. As a consequence, the canoncical momentum of elastic deformations is not equal to the kinetic momentum.

Integrating the kinetic-energy density over the helix volume and substracting the double-well potential $V(q)$ yields the Lagrangian
\begin{equation} 
    L = \frac{M}{2} \int_{-\pi/2}^{\pi/2} \frac{d\varphi}{\pi}\dot{\boldsymbol{\xi}}^2(\varphi) - V(q) = \frac{\mu}{2} \dot{q}^2  + \frac{1}{2} {\boldsymbol \omega} \cdot {\rm I}(q) {\boldsymbol \omega} + \frac{2 M  r}{\pi^2} \dot{q}\, {\boldsymbol \omega} \cdot {\bf n}_1 - V(q),
\end{equation}
with $\mu = M/12$. Comparing this with Eq.\,\eqref{eq:lagrangian} shows that $\kappa = 2 M r \pi^2$. The opening-dependent inertia tensor is
\begin{align}
    {\rm I} = & \frac{Mr^2}{2} \left [ \mathds{1} +  {\bf n}_3 \otimes  {\bf n}_3 - \frac{4 q}{\pi^2 r} \left ( {\bf n}_2 \otimes  {\bf n}_3 +  {\bf n}_3 \otimes  {\bf n}_2 \right )-\frac{8}{\pi^2} \left ( {\bf n}_2 \otimes {\bf n}_2 + {\bf n_3} \otimes {\bf n_3} \right )  + \frac{q^2}{6r^2} \left ( {\bf n}_1 \otimes {\bf n}_1 + {\bf n}_2 \otimes {\bf n}_2\right ) \right ].
\end{align}

The resulting Hamiltonian is of the form \eqref{eq:hamil}. It can be further simplified by neglecting terms proportional to $q^2/r^2$ and using that the effective inertia tensor fulfills
\begin{equation}
    \Lambda {\bf n}_1 = \frac{M r^2}{2}  \left(1 - \frac{96}{\pi^4}\right ) {\bf n}_1,
\end{equation}
yielding
\begin{align}\label{eq:hamil1}
    H = & \frac{1}{2} {\bf J} \cdot \Lambda^{-1}{\bf J} + \frac{p^2}{2 \mu_*}+ V(q)  - \frac{4p}{\pi^2 \mu_* r} \,  {\bf J} \cdot {\bf n}_1,
\end{align}
where we defined the effective mass $\mu_*/\mu = 1 - 96/\pi^4$.

{\it Full helix---} The full helix is parametrized by $\boldsymbol{\xi}(\varphi) = r ( {\bf n}_1 \cos \varphi + {\bf n}_2 \sin \varphi ) + \varphi q {\bf n}_3 / 2\pi$ with $\varphi \in [-\pi,\pi]$,  see Fig.\,\ref{fig:helicenestruct}. All other conventions are as for the half helix. Again, integrating the kinetic-energy density over the helix volume yields
\begin{equation}
    L = \frac{\mu}{2} \dot{q}^2  + \frac{1}{2} {\boldsymbol \omega} \cdot {\rm I}(\Omega,q) {\boldsymbol \omega} + \frac{Mr}{2 \pi} \dot{q}\, {\boldsymbol \omega} \cdot {\bf n}_1 - V(q),
\end{equation}
with $\mu = M/12$ and $\kappa = M r/2 \pi$. The inertia tensor reads
\begin{align}
    {\rm I} = & \frac{Mr^2}{2} \left [ \mathds{1} +  {\bf n}_3 \otimes  {\bf n}_3 - \frac{q}{\pi r} \left ( {\bf n}_2 \otimes  {\bf n}_3 +  {\bf n}_3 \otimes  {\bf n}_2 \right ) + \frac{q^2}{6r^2} \left ( {\bf n}_1 \otimes {\bf n}_1 + {\bf n}_2 \otimes {\bf n}_2\right ) \right ].
\end{align}

The Hamiltonian \eqref{eq:hamil} can be simplified by neglecting terms of order $q^2/r^2$,
\begin{align} \label{eq:hamil2}
    H = & \frac{1}{2} {\bf J} \cdot \Lambda^{-1} {\bf J} + \frac{p^2}{2 \mu_*}+ V(q)  - \frac{p}{\pi \mu_* r} \,  {\bf J} \cdot {\bf n}_1,
\end{align}
with $\mu_*/\mu = 1 - 6/\pi^2$. In Figs. \ref{fig:diffractionpattern} and \ref{fig:deco}, we used the full-helix model to describe the ro-vibrational phase averaging dynamics of [4]-helicene derivatives.

\section{Vibrational two-level approximation} \label{app:twolevel}

Canonical quantization of the Hamiltonian \eqref{eq:hamil} replaces the phase-space coordinates by operators (denoted by sans-serif characters) obeying the canonical commutation relations. For instance, this implies $[{\sf p},{\boldsymbol{\sf{ J}}}] = 0$.

The tunnelling dynamics can be approximately described by the two lowest-lying vibration states, which are well separated from all higher vibration states (see main text). The two vibration states can be expressed as even and odd superpositions of the left- and right-handed states $|L\rangle$ and $|R\rangle$, which are well approximated as harmonic oscillator groundstates of mass $\mu$, frequency $\omega$, and centred at $\pm \ell$,
\begin{equation}\label{eq:hostate}
    \langle q|L\rangle = \left (\frac{\mu \omega}{\pi \hbar} \right )^{1/4} \exp \left [ - \frac{\mu \omega}{2 \hbar} \left (x + \ell\right ) ^2 \right ],
\end{equation}
and likewise for $R$ localized at $\ell$.

Describing the double-well potential by $V(q) = \mu \omega^2\left ( q^2 - \ell^2 \right )^2/8 \ell^2$, with barrier height $\mu \omega^2 \ell^2/8$, the tunnelling splitting of a non-rotating molecule can be calculated as
\begin{align}
    \Delta & = -\frac{1}{\hbar} \left \langle L \left | \frac{p^2}{2\mu} + V(q) \right | R \right \rangle \nonumber \\
& = \frac{\omega}{8} \left ( 3 \zeta - \frac{3}{4 \zeta} -1 \right ) e^{-\zeta}.
\end{align}
Here, we introduced the dimensionless parameter $\zeta = \mu \omega \ell^2 / \hbar$. Since the tunnelling splitting $\Delta$ is defined for non-rotating molecules it depends on the effective mass $\mu$ rather than $\mu_*$.

{\it Half helix---} For a rotating half helix, the effective mass $\mu$ is replaced by $\mu_* = \mu ( 1 - 96/\pi^4)$, so that the corresponding tunnelling splitting becomes
\begin{align}
    \Delta_* & = \Delta - \frac{48}{\hbar \mu_* \pi^4} \langle L|p^2 | R \rangle \nonumber \\
    & = \Delta + \frac{48\omega}{\pi^4}\frac{\mu}{\mu_*} \left (\zeta - \frac{1}{2} \right ) e^{-\zeta}.
\end{align}
For values $\zeta > 1/2$, the tunnelling splitting is larger if rotational degrees of freedom are included than in the pure vibration Hilbert space, $\Delta_* > \Delta$.

The resulting half-helix Hamiltonian reads
\begin{align}
    {\sf H} \simeq & \left ( - \hbar \Delta_* + i \gamma \, {\boldsymbol{\sf J}} \cdot {\bf n}_1 + \frac{1}{2} {\boldsymbol{ \sf J}}\cdot {\sf \Lambda}_{LR}^{-1}{\boldsymbol{\sf J}} \right ) \otimes |L\rangle \langle R| + {\rm h.c.} + {\sf H}_L \otimes |L\rangle \langle L| +  {\sf H}_R \otimes |R\rangle \langle R|,
\end{align}
Here $\gamma$ denotes the Coriolis coupling and $\Lambda_{LR} =  \Lambda(0) /\langle L| R \rangle$ quantifies rotation-induced tunnelling. The centrifugal coupling frequency follows as
\begin{align}
    \gamma = & \frac{4i}{\pi^2 \mu_* r} \langle L | p | R \rangle = \frac{4\mu \omega \ell}{\pi^2 \mu_* r} e^{-\zeta},
\end{align}
and the overlap $\langle L |R \rangle  = \exp( - \zeta)$.

{\it Full helix---} A similar calculation for the full helix shows that 
\begin{align}
    \Delta_* & = \Delta + \frac{3\omega}{\pi^2}\frac{\mu}{\mu_*} \left (\zeta - \frac{1}{2} \right ) e^{-\zeta},
\end{align}
and
\begin{align}
    \gamma = \frac{\mu \omega \ell}{\pi \mu_* r} e^{-\zeta},
\end{align}
with $\mu_*/\mu = 1 - 6/\pi^2$.

\section{Rotation states of left- and right-handed enantiomers}\label{app:rotstates}

{\it Half helix---} The left- and right-handed effective tensors of inertia $\Lambda_{L/R}$ have identical eigenvalues but distinct eigenvectors. Up to linear order of $\ell/r$, the eigenvalues are $\lambda_1 = M r^2 ( 1- 96/\pi^4)/2$, $\lambda_2 = Mr^2(1 - 8/\pi^2)/2$, and $\lambda_3 = M r^2(1  - 4/\pi^2)$ with corresponding eigenvectors
\begin{subequations}\label{eq:vectors}
\begin{align}
    {\bf e}_1^L = {\bf n}_1, & \quad {\bf e}_2^L \simeq {\bf n}_2 - \frac{4\ell}{\pi^2 r} {\bf n_3},  \quad  {\bf e}_3^L \simeq {\bf n}_3 + \frac{4\ell}{\pi^2 r} {\bf n_2}, \\
    {\bf e}_1^R = {\bf n}_1, & \quad  {\bf e}_2^R \simeq {\bf n}_2 + \frac{4\ell}{\pi^2 r} {\bf n_3},  \quad  {\bf e}_3^R \simeq {\bf n}_3 - \frac{4 \ell}{\pi^2 r} {\bf n_2}.
\end{align}
\end{subequations}
The principal axes of the left- and right-handed enantiomers are thus related by a rotation by the angle $8 \ell/\pi^2 r$ around the body-fixed axis ${\bf n}_1$.

The resulting rotational Hamiltonians describe asymmetric rigid rotors with moments of inertia $\lambda_1 < \lambda_2 < \lambda_3$. Again, their eigenvalues are identical but their eigenstates are related by the rotation
\begin{equation}\label{eq:rotation}
    |\eta; R \rangle = \exp \left ( -\frac{i}{\hbar}\frac{8 \ell}{\pi^2 r}{\boldsymbol{\sf J}}\cdot  {\bf n}_1 \right ) \left \vert \eta; L \right \rangle.
\end{equation}
In order to calculate the eigenstates, we numerically diagonalize the Hamiltonian $H_L$ by expanding its eigenstates in the symmetric-top basis $|jkm\rangle$, which are simultaneous eigenstates of the total angular momentum ${\bf J}^2 |jkm\rangle = \hbar^2 j(j+1)|jkm\rangle$, the body-fixed angular momentum ${\bf J}\cdot {\bf n}_1 |jkm\rangle = \hbar k |jkm\rangle$, and the space-fixed angular momentum component ${\bf J}\cdot {\bf e}_x |jkm\rangle = \hbar m |jkm\rangle$. Here, $m,k = -j,\ldots,j$.

The rotational Hamiltonian ${\sf H} = {\sf L}_1^2/2 \lambda_1 + {\sf L}_2^2/2 \lambda_2 + {\sf L}_3^2/2\lambda_3$, with ${\sf L}_i = \boldsymbol{\sf J}\cdot {\bf n}_i$ the body-fixed angular momentum operators projected on the eigenvectors \eqref{eq:vectors}, commutes with ${\sf L}^2 = {\sf L}_1^2 + {\sf L}_2^2+{\sf L}_3^2$ and with ${\bf J} \cdot {\bf e}_x$ so that $j$ and $m$ are good quantum numbers. The remaining $2j+1$ eigenstates for fixed $j$ and $m$ are labeled by $\eta = (jnm)$ with $n = -j, \ldots ,j$, so that
\begin{equation}
    |\eta; L\rangle = \sum_{k = -j}^j c_{k}^\eta \,|jkm\rangle.
\end{equation}
The resulting eigenvalue problem for the coefficients $c_k^\eta$ is found by using
\begin{subequations}
\begin{align}
    \langle jkm |L_2^2 | jkm \rangle & = \langle jkm |L_3^2 | jkm \rangle = \frac{\hbar^2}{2} [ j(j+1) - k^2], \\
    \langle jkm |L_2^2 | jk+2m \rangle & = \langle jk+2m |L_2^2 | jkm \rangle  = \frac{\hbar^2}{4} \sqrt{(j-k)(j-k-1)(j+k+1)(j+k+2)}, \\
    \langle jkm |L_3^2 | jk+2m \rangle & = \langle jk+2m |L_3^2 | jkm \rangle = -\frac{\hbar^2}{4} \sqrt{(j-k)(j-k-1)(j+k+1)(j+k+2)},
\end{align}
\end{subequations}
and using that all other matrix elements are zero. Given the coefficients $c_k^\eta$, the overlaps follow from Eq.\,\eqref{eq:rotation},
\begin{equation}
    \langle \eta;L|\eta; R\rangle = \sum_{k = -j}^j |c_k^\eta|^2 \exp \left ( -i k \frac{8 \ell}{\pi^2 r} \right ).
\end{equation}

{\it Full helix---} In a similar fashion, the eigenvalues of the effective inertia tensor of the left- and right-handed full helix are $\lambda_1 = M r^2 ( 1- 6/\pi^2)/2$, $\lambda_2 = Mr^2/2$, and $\lambda_3 = M r^2$ with corresponding eigenvectors
\begin{align}
    {\bf e}_1^L = {\bf n}_1, & \quad {\bf e}_2^L \simeq {\bf n}_2 - \frac{\ell}{\pi r} {\bf n_3},  \quad  {\bf e}_3^L \simeq {\bf n}_3 - \frac{\ell}{\pi r} {\bf n_2}, \\
    {\bf e}_1^R = {\bf n}_1, & \quad  {\bf e}_2^R \simeq {\bf n}_2 + \frac{\ell}{\pi r} {\bf n_3},  \quad  {\bf e}_3^R \simeq {\bf n}_3 + \frac{\ell}{\pi r} {\bf n_2}.
\end{align}
The left- and right-eigenvectors are thus related by a rotation by the angle $2 \ell/\pi r$ around the body-fixed axis ${\bf n}_1$, so that the eigenstates of the rotational Hamiltonians fulfill
\begin{equation}
    |\eta; R \rangle = \exp \left ( -\frac{i}{\hbar}\frac{2 \ell}{\pi r}{\boldsymbol{\sf J}}\cdot  {\bf n}_1 \right ) \left \vert \eta; L \right \rangle,
\end{equation}
and
\begin{equation}
    \langle \eta;L|\eta; R\rangle = \sum_{k = -j}^j |c_k^\eta|^2 \exp \left ( -i k \frac{2 \ell}{\pi R} \right ).
\end{equation}

Finally, we remark that the Hamiltonian ${\sf H}_{L/R}$ is spatially isotropic and thus its eigenenergies $E_\eta$ are independent of the space-fixed angular momentum quantum number $m$, implying that the orientational diagonal elements of the ro-vibrational state \eqref{eq:rho} fulfill $\langle \Omega | \rho | \Omega \rangle = {\rm tr}_{\rm rot}(\rho)/8 \pi^2$. Here, ${\rm tr}_{\rm rot}(\cdot)$ denotes the partial trace over the rotational Hilbert space. This follows from the completeness of the Wigner $D$-matrices \cite{edmonds1996}
\begin{equation}
    \sum_{m = -j}^j \langle \Omega|jkm\rangle \langle j k' m|\Omega\rangle = \frac{2j+1}{8 \pi^2} \delta_{kk'}.
\end{equation}
This relation expresses that all orientations are equally probable.

\section{Far-field diffraction}\label{app:ffdiff}
Writing the phase \eqref{eq:gratephase} as $\phi \cos^2(2 \pi \sin \theta x/\lambda)$ and the phase \eqref{eq:helphase} as $\chi_{L/R} \sin(4 \pi \sin \theta x/\lambda)$, the grating transformation \eqref{eq:grattrafo} for fixed handedness takes on the form
\begin{equation} \label{eq:trafo}
    t^{L}(x) = \sum_{n = -\infty}^\infty i^n J_n \left ( \frac{\phi}{2} + \chi_L \right ) \exp \left (i \frac{4 \pi n \sin \theta x}{\lambda} \right ),
\end{equation}
and likewise for $R$. Here, we used that $\vartheta = \pi/2$. The resulting far-field diffraction pattern follows from propagating the spatial wave-function $|x_0\rangle$ first to the grating with the unitary ${\sf U}_1$, and then to the detection screen at distance $d$ behind the grating,
\begin{align}
    \langle x|\Phi_L\rangle \sim & \sum_{n = -\infty}^\infty i^n J_n \left ( \frac{\phi}{2} + \chi_L \right ) \int_{-w/2}^{w/2} dx' \left \langle x' \left | {\sf U}_1 \right |  x_0 \right \rangle  \exp \left [ -i \frac{m v_z x'}{\hbar d} \left ( x - n\frac{4 \pi \sin \theta \hbar d}{m v_z \lambda} \right ) \right ].
\end{align}
The spacing of neighboring diffraction orders is $4 \pi\sin \theta  \hbar d/mv_z \lambda$.

\section{Interferometric sensing and motional decoupling}\label{app:sensing}

Filtering the $|R\rangle$-state at the grating and imprinting the relative phase $\varphi$ at time $t$ yields
\begin{equation}
    |\Psi_{\eta R}'\rangle = \cos( \Delta_\eta t) |\eta R\rangle +i e^{i \gamma_\eta} e^{i \varphi} \sin(\Delta_\eta t) |\eta L\rangle,
\end{equation}
which after a free evolution for time $t'$ gives $|\Psi_{\eta R}''\rangle = a_\eta(t,t';\varphi)  |\eta R\rangle +i e^{i \gamma_\eta}  b_\eta(t,t';\varphi) |\eta L\rangle$ with the coefficients
\begin{subequations}
\begin{align}
    a_\eta(t,t';\varphi) = & \cos( \Delta_\eta t) \cos( \Delta_\eta t') - e^{i \varphi} \sin( \Delta_\eta t) \sin( \Delta_\eta t'), \\
    b_\eta(t,t';\varphi) = & \cos( \Delta_\eta t) \sin( \Delta_\eta t')  + e^{i \varphi} \sin( \Delta_\eta t) \cos( \Delta_\eta t').
\end{align}
\end{subequations}
The probability $\overline{P}_R(t,t') = \sum_\eta p_\eta |a_\eta(t,t';\varphi)|^2$ at fixed times $t$ and $t'$ strongly depends on the relative phase $\varphi$ and can thus be used to measure it. In particular, a straight forward calculation yields
\begin{align}
    \overline{P}_R(t,t') = &  \cos^2\left ( \frac{\varphi}{2}\right ) P_R(t + t') + \sin^2\left ( \frac{\varphi}{2}\right ) P_R(t - t'),
\end{align}
where $P_R(t)$ denotes the probability of measuring the $R$-enantiomer in the absence of the phase shift. Thus, for $t = t'$, the unperturbed signal is modulated with $\cos^2(\varphi/2)$. In addition, applying $\varphi = \pi$ implements motional decoupling by undoing the rotational phase averaging. For $t' = t$, this yields the initial condition $a_\eta(t,t;\pi) = 1$ and $b_\eta(t,t;\pi) = 0$, as illustrated in Fig.\,\ref{fig_supp:lasergratings}(a).

\begin{figure}[t!]
    \centering
    \includegraphics[width = 0.9 \textwidth]{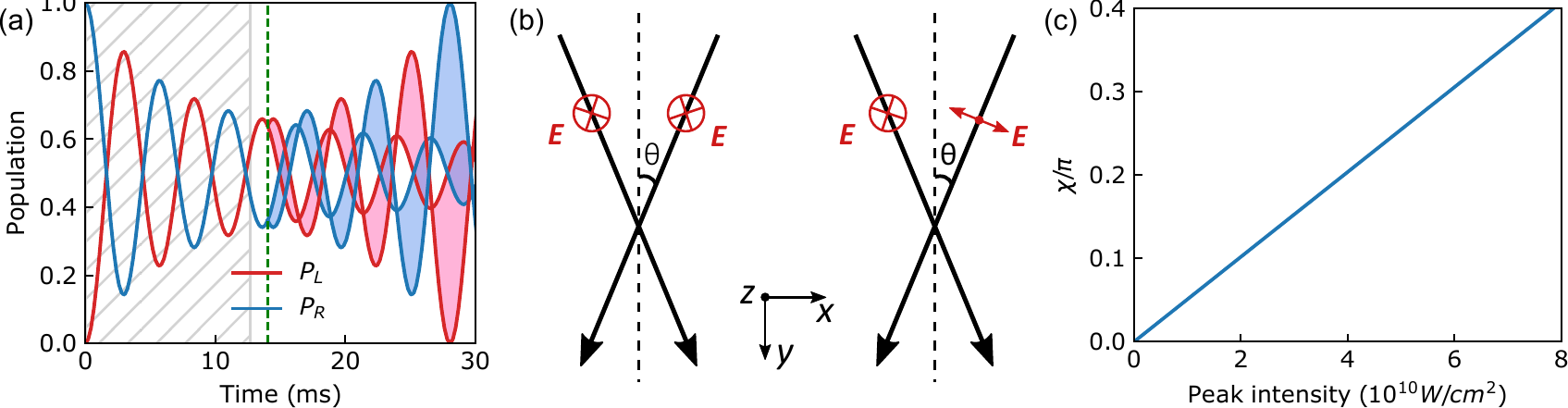}
    \caption{(a) Motional decoupling implemented by applying a relative phase shift $\phi=\pi$ at the time indicated by the dashed green line, which effectively reverses the rotational phase averaging of the tunnelling signal. (b) Optical gratings created by superposing two linearly polarized laser fields each of electric field amplitude $E$ and copropagating in the transverse $x-y$ plane. Left: for the standing-wave grating, the two laser beams have the same polarization. Right: for the helicity grating, the two beams have orthogonal polarizations. (c) Phase shift of the helicity grating as a function of peak optical intensity in each laser beam considering Gaussian-shaped pulse envelope and beam profile.}
    \label{fig_supp:lasergratings}
\end{figure}

\section{Decoherence dynamics}\label{app:deco}

The ro-vibrational dynamics in presence of decoherence with the rotation-state-independent rate $\Gamma_{LR}$ is described by the Lindblad master equation
\begin{align}\label{eq:master}
    \partial_t \rho = -&  \frac{i}{\hbar} [H,\rho]  + \Gamma_{LR} \sum_\eta \left ( \sum_{h = L,R} |\eta h\rangle \langle \eta h|\rho |\eta h\rangle \langle \eta h| - \rho \right ),
\end{align}
with the tunnelling Hamiltonian \eqref{eq:rovibham}. The rate $\Gamma_{LR}$ is given by \eqref{eq:decorate} if the molecule is inside the collision chamber, and zero before and after it. Since this Hamiltonian induces no rotational coherences and since the initial rotation state is thermal with weights $p_\eta$, we can write the ro-vibration state at all times as a mixture of vibration states, $\rho = \sum_\eta p_\eta \rho_\eta$.

The dynamics of each vibration state $\rho_\eta$ with ${\rm tr}(\rho_\eta) = 1$ can be calculated by recasting the master equation \eqref{eq:master} for each $\eta$ in the form $\partial_t {\bf c}_\eta = i {\sf A}_\eta {\bf c}_\eta$, with the vector of matrix elements ${\bf c}_\eta = (\langle L|\rho_\eta |L\rangle,\langle R|\rho_\eta |R\rangle,\langle L|\rho_\eta |R\rangle,\langle R|\rho_\eta |L\rangle)^T$ and the non-Hermitian matrix
\begin{equation}
    {\sf A}_\eta = \left ( \begin{array}{cccc}
    0 & 0 & - \Delta_\eta e^{-i \gamma_\eta}&  \Delta_\eta e^{i \gamma_\eta} \\
    0 & 0 & \Delta_\eta e^{-i \gamma_\eta} & - \Delta_\eta e^{i \gamma_\eta} \\
    -\Delta_\eta e^{i \gamma_\eta} &  \Delta_\eta e^{i \gamma_\eta} & i\Gamma_{LR}& 0 \\
     \Delta_\eta e^{-i \gamma_\eta} & - \Delta_\eta e^{-i \gamma_\eta} & 0& i\Gamma_{LR}
    \end{array}
    \right ).
\end{equation}
The solution to this equation is given by ${\bf c}_\eta(t) = \exp ( i {\sf A}_\eta t) {\bf c}_\eta(0)$.

Filtering $R$-states at the grating, ${\bf c}_\eta(0) = (0,1,0,0)^T$, and denoting by $T$ the time from the grating to the screen, the state at the screen is
\begin{equation}
   {\bf c}_\eta(T) = \left ( \begin{array}{c}
   \sin^2 (\Delta_\eta T) \\
   \cos^2(\Delta_\eta T) \\
   i e^{i\gamma_\eta} \sin(2 \Delta_\eta T)/2 \\
   -i e^{-i\gamma_\eta} \sin(2 \Delta_\eta T) /2
   \end{array} \right ),
\end{equation}
since there is no notable decoherence in the interferometer. After traversing the collision chamber for time $\tau$, the probability of measuring the $R$ state can be calculated as
\begin{equation} \label{eq:PLdeco}
    P_{\eta R}(\tau) = \frac{1}{2} + \frac{1}{2}e^{-\Gamma_{LR}\tau/2} \left [ \cos(2 \omega_\eta \tau)\cos(2 \Delta_\eta T) - \frac{\Delta_\eta}{\omega_\eta} \sin(2 \omega_\eta \tau) \sin (2 \Delta_\eta T) + \frac{\Gamma_{LR}}{4 \omega_\eta} \sin(2 \omega_\eta \tau)\cos(2 \Delta_\eta T) \right ],
\end{equation}
with $\omega_\eta = \sqrt{\Delta_\eta^2 - (\Gamma_{LR}/4)^2}$. The final signal is obtained by summing over all rotation states with weights $p_\eta$,
\begin{equation}
    P_R(\tau) = \sum_\eta p_\eta P_{\eta R}(\tau).
\end{equation}
Equation \eqref{eq:PLdeco} contains two relevant limiting cases: (i) For vanishing decoherence $\Gamma_{LR} \ll \Delta_\eta$, the probability of measuring the $R$-chiral state reduces to $P_{\eta R}(\tau) \simeq 1/2 + \exp ( - \Gamma_{LR} \tau/2) \cos[2\Delta_\eta ( T + \tau)]/2$. (ii) For long tunnelling times $\Delta_\eta \gg \Gamma_{LR}$ and $\Delta_\eta T \ll 1$, the state stays pure and there is no influence of decoherence, $P_{\eta R} \simeq 1$.

\section{Choice of molecules and computational methods} \label{app:methods}

The possibility that molecular chirality is induced in overcrowded structures of angularly fused benzene rings was noted for 4,5-dimethyl phenanthrene in the 1940s \cite{newman1940,newman1948}. It was demonstrated by synthesis and enantioresolution of [6]-helicene with subsequent measurement of optical rotation \cite{newman1956}, but it took four decades to demonstrate it
experimentally for the original 4,5-dimethyl phenanthrene \cite{armstrong1987}, due to fast racemisation. For the present work, fast racemisation and comparatively low barrier of stereomutation is, however, desirable to obtain tunneling splittings in the range of $100~\mathrm{Hz}$ to $50~\mathrm{kHz}$. 

Tunneling splittings of $150~\mathrm{kHz}$ have been observed experimentally for the axially chiral inorganic compound disulfane (H$_2$S$_2$) \cite{winnewisser1991} and splittings about an order or magnitude smaller have been predicted for TSOT, H$_2$Se$_2$ and H$_2$Te$_2$ \cite{quack2003,gottselig2003,gottselig2004,sahu2020}, but the optical rotation for this compound class is expected to be small. Helicenes and derivatives thereof, in contrast, are well-known for comparatively large optical rotation
\cite{newman1956,fitts1955,brown1971,brown1971a,brickell1971,martin1974,grimme1996,buss1996,nakai2012,gingras2013a},
which is crucial to realise the grating transformation in the present work. Optical rotation increases with increasing $n$ in [$n$]-helicenes \cite{martin1974} whereas activation energies for racemisation become lower for smaller helicenes but reach a plateau for $n>6$
\cite{martin1972,martin1974,martin1974a,lindner1975,grimme1996,janke1996}.

In [4]-helicene (also denoted as 3,4-benzphenanthrene or
benzo[c]phenanthrene; see Fig.\,\ref{fig:helicenestruct}), the barrier for racemisation is reported to be too low to allow separation of enantiomers at room temperature \cite{newman1952}. Lindner was the first to study the racemisation mechanism of helicenes computationally and found a planar transition structure for [4]-helicene with an activation barrier of about $50\,\mathrm{THz}$, which was confirmed also in later studies \cite{grimme1996,barroso2018}, which was the motivation for choosing [4]-helicenes and derivatives thereof as a starting point in the present work. 

Tunability of the activation barrier for stereomutation by substitution is well known for [4-6]-helicenes \cite{newman1952,janke1996,ravat2017} and Ref.~\cite{ravat2017} found a correlation between the torsional twist in the equilibrium structure induced by substituents in the fjord region of [5]-helicenes and the corresponding activation barrier for stereomutation,
by which a quick assessment of expected barrier heights is possible. Additional fine-tuning of the tunneling splittings is possible by isotopic substitution or by replacement of hydrogen with fluorine on the periphery because of the exponential dependence on the square root of the tunneling
mass \cite{gamow1928,berger2004}. Our primary target is thus to identify [4]-helicene derivatives that can bracket the desired tunneling splitting, while maintaining the comparatively large optical rotation that is characteristic for the helicene family. Chemical tuning
will allow to determine subsequently ideal compounds.

The equilibrium structures of [4]-helicene as determined in
Ref.~\cite{nakai2012} on the density functional theory (DFT) level B97-D with the D2 dispersion correction and a basis set of triple zeta quality (TZVP) were taken as input structures. Equilibrium and transition structures were in the present work then determined on the same level of theory with the program package Turbomole \cite{turbomole}.

The isotropic part of the optical rotatory dispersion tensor
$\boldsymbol\beta(\omega)$ at real and imaginary frequencies and the frequency dependent electric dipole polarisability tensor $\boldsymbol\alpha(\omega)$ were calculated with the same program package on the time-dependent DFT level with the B3LYP hybrid density functional and a triple-zeta basis set (TZVPP) basis set and the recommended basis sets for the resolution of the identity (RI) of the Coulomb term (RI-J). $\boldsymbol\beta(\omega)$ is related to the gyration tensor $\boldsymbol G'(\omega)$ \cite{barron2009} by
\begin{equation}
\boldsymbol G'(\omega) = - \omega \,\boldsymbol \beta(\omega)
\end{equation}
and can possess in contrast to $\boldsymbol G'(\omega)$ a finite static limit \cite{amos1982}. The proper units that result when $G'$ and $\beta$ are reported in atomic units, are obtained by the collection of the following constants
\begin{align}
[G'] &= e a_0  \, e \hbar m_\mathrm{e}^{-1}\, E_\mathrm{h}^{-1}
      = e^2 a_0^3 \hbar^{-1} \\
[\beta] &= e a_0 \, e \hbar m_\mathrm{e}^{-1}\, \hbar E_\mathrm{h}^{-2}
         = e^2 a_0^5 m_\mathrm{e} \hbar^{-2},
\end{align}
with $e$ being the electric charge of the proton, $a_0$ being the Bohr radius, and $E_\mathrm{h}$ the Hartree energy.

For the equilibrium structure of [4]-helicene, we estimate the static-limit isotropic optical rotatory dispersion $\beta_\mathrm{iso}\approx 4\,e^2 a_0^5 m_e \hbar^{-2}$, which is only a factor 5 to 6 smaller than the corresponding value in [6]-helicene and thus consistent with the ratio computed in Ref.~\cite{nakai2012} for the frequency dependent optical rotation of the two compounds at the sodium D-line.

For the equilibrium structure of the naphthindole (see Fig.\,\ref{fig:helicenestruct}), we obtain
$\beta_\mathrm{iso} \approx 1.39 ~e^2 a_0^5 m_e \hbar^{-2}$, which is nearly a factor of 3 smaller than the value for [4]-helicene. The lowest harmonic vibrational frequency remains almost unchanged (about $2~\mathrm{THz}$), but the electronic barrier for stereomutation drops to only $7\,\mathrm{THz}$.

Tunnelling splittings were estimated with the instanton code as implemented in the program package Molpro \cite{molpro} on the DFT level with the B3LYP functional and the 3-21G basis set for all atoms. For further details on the instanton approach we refer to Ref.\,\cite{garg2000}. The computed action $S$ for the instanton path of [4]-helicene is $\sim 55\hbar$ using 160 beads and $\beta=20000\,\hbar$. This gives a tunneling splitting of $\sim 12$\,pHz. Using the B97-D functional with the D4 dispersion correction and a correlation consistent double zeta basis set (cc-pVDZ) for all atoms, instanton paths with up to 65 beads were calculated. The resulting action is $\sim 52\,\hbar$. As this term enters exponentially in the calculation of the tunneling splitting with $\Delta E_\pm \propto \exp(−S/\hbar)$, it essentially determines the magnitude of the tunneling splitting.  This gives a difference of 1-2 orders of magnitude in the tunneling splitting for the different methods, which is sufficient for our estimates.

We thus considered several other derivatives and turned to the naphthindole sketched in Fig.\,\ref{fig:helicenestruct}. Although the instanton approach becomes less accurate for shallow tunneling situations \cite{amos1982}, it can still provide an order of magnitude for the tunneling splitting. With the  B3LYP functional we calculated 320 beads with $\beta=15000\,\hbar$ for naphthindole, which gave an action $\sim 14\,\hbar$ and a splitting $\sim 40$\,MHz. At 160 beads the action has already converged to $(14.25\pm 0.1)\,\hbar$ for $\beta$ values between $15000\,\hbar$ and $25000\,\hbar$, giving a sufficient accurate estimate of the splitting. With the better B97-D functional using 65 beads and the same $\beta$ the action is $\,12\,\hbar$. Though the action has not converged in these calculations, we can estimate of the splitting in the range of 1 to 200\,MHz, so that the two compounds selected here clearly bracket the desired range of 100\,Hz to 50\,kHz used in the present estimates. 

\section{Simulation parameters}\label{app:experiment}

For the simulations presented in Figs. \ref{fig:diffractionpattern} and \ref{fig:deco} , we consider a molecular beam generated by seeding [4]-helicenes into a supersonic jet of Xenon, yielding a mean velocity $v_z\simeq 315$\,m/s and a ro-vibrational temperature of approximately $4$\,K. The beam is first filtered by a source skimmer with a diameter of $12\,\mu$m, before propagating freely for a distance of $3$\,m to a collimation slit with an opening of $6\,\mu$m. After passing the slit, the molecules are illuminated by two pulsed optical gratings. The gratings are realized by superposing two laser beams of the same intensity and co-propagating at an angle $2\theta = 1$ perpendicular to the molecular beam, as illustrated in Fig.\,\ref{fig_supp:lasergratings}. For the first grating, the two laser beams have their polarization aligned parallel to each other, see Fig.\,\ref{fig_supp:lasergratings}(b). The interference leads to an intensity gradient in the transverse direction $x$ and gives rise to an electric-dipole interaction potential in the form of
\begin{equation}
V_\eta^L(x,t) = -\alpha_\eta^L E^2(t)\left[1+\cos{(\zeta x)}\right]\,.
\end{equation}
Integrating this potential over the temporal profile of the laser pulse yields the phase shift (\ref{eq:gratephase}). The right-handed enantiomer experiences the same phase shift since $\alpha^L_\eta=\alpha^R_\eta$.

The second optical grating, acting directly after the first, so that the motion of the molecules between the gratings is negligible, is created in a similar fashion, but with the polarization of the two beams perpendicular to each other. Here, the superposition of the two laser beams creates an optical helicity-density gradient which interacts with the molecules via their electric-magnetic dipole polarizability pseudotensor $G_{\eta}^{L/R}$ \cite{Cameron2014a, Cameron2014b}. The resulting interaction potential can be written as
\begin{equation}
\tilde{V}^L_\eta(x,t)=-\frac{G^L_\eta}{c}E_{\rm hel}^2(t)\cos^2{\theta}\sin{\left(\zeta x+\vartheta\right)}\,,
\end{equation}
yielding the phase shift (\ref{eq:helphase}). The right-handed enantiomer experiences the opposite phase shift since $G^L_\eta= - G^R_\eta$. After traversing the gratings, the molecule propagates for a distance of $4$\,m, where a screen with a slit of $10\,\mu$m-opening blocks all but a single diffraction peak. The requirement on the total length of the interferometer can be alleviated by using a slower molecule beam, e.g., from a buffer gas cell \cite{hutzler2012}. 

The two optical gratings can be generated by a pulsed Nd:YAG laser operating at 1064\,nm with pulse duration of 10\,ns. Both gratings can be implemented by splitting the laser beam  with a beam splitter and superposing them in the interaction region using a Mach-Zehnder configuration \cite{mohanty2005,cipparrone2010}. Considering a collimated Gaussian beam profile with beam diameter $2 w_{x,y}=0.5$\,mm, pulse energy of 250\,mJ is necessary in each beam to reach a phase shift of $\pi/10$ for the helicity grating. Generating the same phase shift with the standing-wave grating, requires only $1.25\,\mu$J in each beam. The phase shifts in Eq.\,\ref{eq:gratephase} and Eq.\,\ref{eq:helphase} assume a flat intensity profiles in the interaction region of $6\,\mu$m, which can be achieved by using beam shaping optics or by adapting an active resonator design of the Nd:YAG laser. The relation between the phase shift of the helicity grating versus the peak optical intensity of each laser beam is plotted in Fig.\,\ref{fig_supp:lasergratings}(c). To reach a phase shift of $\pi/10$, a peak intensity of $2\times 10^{10} {\rm W/cm^2}$ is required in each beam. The laser wavelength 1064\,nm is far detuned from the electronic (<400\,${\rm nm}$) and vibrational (>3\,$\mu$m) absorption bands of helicene derivatives. In addition, the absorption of a photon in the grating transfers a momentum kick of magnitude $2\pi\hbar/\lambda$ to the molecule in the propagation direction of the laser beam, so that the corresponding molecules will be filtered out at the screen. The intensity required for the helicity grating is one order of magnitude lower than those employed in sensitive dipole force measurements using aromatic hydrocarbons and Nd:YAG laser at the same wavelength \cite{chung2001, fulton2004}.

To simulate the tunnelling dynamics of [4]-helicene derivatives , we use the full-helix model with $\mu = 24\,$u, $\omega/2 \pi = 3\,$THz, $r = 2.88$\,\AA\;and $\ell = 0.7$\,\AA\;, yielding that Coriolis coupling and the overlaps are negligibly small, while the tunnelling splitting is on the order of $\Delta_*/2\pi \simeq 10^2$\,Hz. To calculate the decoherence dynamics illustrated in Fig.\,\ref{fig:deco}, we considered nitrogen molecules with mass of 28\,u and rotation averaged-polarizability $\alpha_{\rm g}/4 \pi \varepsilon_0 = 1.7$\,\AA${}^3$\;taken from Ref.\,\cite{spelsberg1994}.
\end{widetext}

\end{document}